\documentclass[longbibliography,letterpaper,english,british,reprint, aps]{revtex4-1}
\usepackage[T1]{fontenc}
\usepackage[latin9]{inputenc}
\usepackage{bm}
\usepackage{amsmath}
\usepackage{amssymb}
\usepackage{graphicx}

\makeatletter

\pdfpageheight\paperheight
\pdfpagewidth\paperwidth

\newcommand{\lyxmathsym}[1]{\ifmmode\begingroup\def\b@ld{bold}
  \text{\ifx\math@version\b@ld\bfseries\fi#1}\endgroup\else#1\fi}

\providecommand{\tabularnewline}{\\}

\makeatother

\usepackage{babel}
\begin{document}

\preprint{APS/123-QED}

\title{Tackling solvent effect by coupling electronic and molecular Density
Functional Theory}

\author{Guillaume Jeanmairet}

\affiliation{Sorbonne Université, CNRS, Physico-Chimie des Électrolytes et Nanosystèmes
Interfaciaux, PHENIX, F-75005 Paris, France.}

\affiliation{Réseau sur le Stockage Électrochimique de l'Énergie (RS2E), FR CNRS
3459, 80039 Amiens Cedex, France}
\email{guillaume.jeanmairet@sorbonne-universite.fr}

\author{Maximilien Levesque}

\affiliation{PASTEUR, Département de chimie, École normale supérieure, PSL University,
Sorbonne Université, CNRS, 75005 Paris, France}

\author{Daniel Borgis}

\affiliation{PASTEUR, Département de chimie, École normale supérieure, PSL University,
Sorbonne Université, CNRS, 75005 Paris, France}

\affiliation{Maison de la Simulation, CEA, CNRS, Université Paris-Sud, UVSQ, Université
Paris- Saclay, 91191 Gif-sur-Yvette, France}
\begin{abstract}
Solvation effect might have a tremendous influence on chemical reactions.
However, precise quantum chemistry calculations are most often done
either in vacuum neglecting the role of the solvent or using continuum
solvent model ignoring its molecular nature. We propose a new method
coupling a quantum description of the solute using electronic density
functional theory with a classical grand-canonical treatment of the
solvent using molecular density functional theory. Unlike previous
work, both densities are minimised self consistently, accounting for
mutual polarisation of the molecular solvent and the solute. The electrostatic
interaction is accounted using the full electron density of the solute
rather than fitted point charges. The introduced methodology represents
a good compromise between the two main strategies to tackle solvation
effect in quantum calculation. It is computationally more effective
than a direct quantum-mechanics/molecular mechanics coupling, requiring
the exploration of many solvent configurations. Compared to continuum
methods it retains the full molecular-level description of the solvent.
We validate this new framework onto two usual benchmark systems: a
water solvated in water and the symmetrical nucleophilic substitution
between chloromethane and chloride in water. The prediction for the
free energy profiles are not yet fully quantitative compared to experimental
data but the most important features are qualitatively recovered.
The method provides a detailed molecular picture of the evolution
of the solvent structure along the reaction pathway. 
\end{abstract}
\maketitle

\section{Introduction\label{sec:Introduction}}

The solvent is often described as the media in which a chemical reaction
between dissolved species, called solutes, takes place. However, it
is well known that besides its role of bringing the reactants in contact,
the solvent has a tremendous influence on the chemical reaction as
it impacts the kinetics and the thermodynamics of the reaction. Organic
chemists have taken advantage of these solvent effects since decades.
For instance, in the 30's, Hughes and Ingold already discussed a theory
of solvation effect for nucleophilic substitution ($\text{S}_{\text{N}}$)
\citep{hughes_55._1935}. In this paper, they reviewed the already
substantial experimental work on the influence of the choice of solvent
on the $\text{S}_{\text{N}}$ reaction and they proposed a theoretical
model accounting for solvent effect. Hughes and Ingold model is based
on a simple hypothesis: only electrostatic interactions are considered.
By examining the stabilising and/or destabilising role of the solvent
on the reactants, products and transition states of the reaction they
were able to rationalise effect of solvent polarity on reaction rates.
However, solvent molecules may also be directly involved in the reaction
mechanism. For instance, Liu et al have shown that adding a single
methanol molecule largely promotes the $\text{S}_{\text{N}}$ reaction
with respect to elimination reaction \citep{liu_how_2018}. Direct
involvement of solvent molecules cannot be captured by macroscopic
consideration as the ones used in Hughes and Ingold model. A good
alternative is to resort to molecular simulation. Since chemical reaction
involves formation and/or breaking of chemical bonds it makes the
use of classical force field difficult: even if reactive force field
exist, their parametrisation are often costly and difficult \citep{han_development_2016,senftle_reaxff_2016}.
Thus, to study chemical reaction the first approach is often to use
a Quantum Mechanics (QM) based method such as electronic density functional
theory (eDFT), Hartree Fock or more advanced technique such as Moller-Plesset
or Coupled Cluster. Due to computational cost, such calculations are
often run in vacuum and at 0~K which neglect solvent effect.

To incorporate solvent effect, the most natural choice is to explicitly
include solvent molecules into the simulation. This is extremely costly
since it increases considerably the number of electrons with respect
to in-vacuo calculations. The finite temperature is even more problematic
since the meaningful quantity is no longer the ground state energy
but the free energy. This means that the calculation should take place
in a statistical ensemble and that a long enough trajectory should
be produced to compute ensemble average with good statistics \citep{frenkel_understanding_2002}.
This is the typical setup of \textit{ab-initio} Molecular Dynamics
(AIMD) calculations \citep{silvestrelli_water_1999}. The problem
of free energy calculation is particularly difficult since it cannot
be written as a (grand) canonical average over the phase space \citep{frenkel_understanding_2002}.
For those two reasons AIMD simulations are limited to efficient QM
techniques. There are almost always based on eDFT and only small systems
can be investigated. 

To circumvent the computational cost of AIMD a natural choice is to
split the studied system into 2 parts. A first one, which is considered
as essential for the description of the chemistry is treated at the
QM level. A second one, which often includes most of the solvent molecules,
for which interactions are described by molecular mechanics (MM) using
classical force fields. This is the well known QM/MM approach which
has been used successfully to tackle a wide variety of systems \citep{lin_qm/mm:_2007}.
The MM part is most often dealt with Molecular Dynamics (MD) or Monte
Carlo (MC). While QM/MM makes it possible to considerably reduce the
numerical cost of evaluating forces as compared to AIMD, the problem
of computing free energy remains. A further simplification is to average
out the solvent degrees of freedom to reduce tremendously the dimensionality
of the problem and the computational cost. This is the strategy adopted
in continuum solvent models (CSM) approaches such as the polarisable
continuum model (PCM) \citep{tomasi_molecular_1994,tomasi_quantum_2005,klamt_cosmo_2011}
where the solvent is described by a dielectric continuum. CSM have
been applied with success but it suffers from several drawbacks. First,
it contains several ah-hoc parameters that are not physically based
but are rather optimised on reference calculations. Second, it completely
loses the description of the solvent at the molecular level. 

Another way to tackle solvation problem is to conserve the QM/MM partition
of the system but to use liquid state theories to deal with the MM
part. Liquid state theories have proven efficient to describe simple
liquids such as hard-sphere or Lennard-Jones fluids \citep{hansen_theory_2006}.
Several approaches such as integral equation theory, its interaction
site approximation (RISM) \citep{chandler_optimized_1972} or classical
density functional theory (cDFT) \citep{ebner_density-functional_1976}
have been applied in that context. The common objective of all these
techniques is to find the equilibrium number solvent density $n$($\bm{r}$),
or equivalently its total correlation function $h(\bm{r})$. For simple
liquid such as hard-sphere or Lennard-Jones fluid these fields solely
depends on the position of the solvent $\bm{r}$. A key quantity entering
all these theories is the direct correlation function between solvent
molecules $c(r)$. 

More recent development have made possible to tackle realistic molecular
fluids such as water or acetonitrile, either based on the integral
equation theory \citep{belloni_efficient_2014} or the molecular density
functional theory (MDFT) \citep{ramirez_density_2002,zhao_molecular_2011,jeanmairet_molecular_2013,jeanmairet_molecular_2013-1,jeanmairet_molecular_2015,jeanmairet_molecular_2016,jeanmairet_study_2019,jeanmairet_molecular_2019}.
In a molecular solute-solvent system the density field $\rho(\bm{r},\bm{\Omega})$
no longer depends solely on the position $\bm{r}$ but also on the
orientation $\bm{\Omega}$ of the solvent. Consequently the direct
correlation function $c(\bm{r},\bm{\Omega}_{1},\bm{\Omega}_{2})$
becomes extremely complicated as it now depends on a position vector
and two orientations.

The RISM approach and its 3D-RISM \citep{kovalenko_three-dimensional_1998}
generalisation circumvent this problem by averaging out the solvent
orientation. The complicated molecular correlation function $c$ is
replaced by simpler 1D site-site correlation functions. The gain in
efficiency is obvious but it is at the price of working with site-site
OZ equation and site-site correlation functions which are no longer
based on a proper statistical physics derivation. Another approach
based on molecular density functional theory (MDFT) has ben proposed
recently to solve the MOZ equation, with the hypernetted chain (HNC)
closure, without making such approximations. The use of projections
onto rotational invariants allows to handle the numerically costly
angular convolution products \citep{ding_efficient_2017} making the
method practical.

Liquid state theories represent a good compromise between CSM and
MD based approaches since it conserves the description of solvent
as made of molecular entities while not requiring the tedious statistical
sampling of solvent degrees of freedom. It is thus natural to propose
a formalism where solutes are described using QM and solvent using
liquid state theories. Since the original paper \citep{ten-no_hybrid_1993},
numerous studies using RISM \citep{naka_rism-scf_1999} or 3D-RISM
\citep{sato_self-consistent_2000,kasahara_solvation_2018,kaminski_modeling_2010,zhou_multi-scale_2011}
have been done to study solvation effect on a solute described by
QM. This method is implemented in widely distributed quantum codes
such as ADF \citep{gusarov_self-consistent_2006,casanova_evaluation_2007}.
Because the development of classical DFT techniques to study molecular
liquids are more recent, less attempt have been done to describe solvation
of QM solutes with this method. The Arias group have developed the
joint DFT framework \citep{petrosyan_joint_2005,petrosyan_joint_2007}
and released the code JDFTx \citep{sundararaman_jdftx_2017} where
the method is implemented. The classical DFT fluids that are implemented
in JDFTx are using functional based on model molecular Hamiltonian
that are parametrised to reproduce some feature of the liquid such
as its non-local dielectric response on external electric field \textemdash an
improvement over PCM. However the performance of this simplified DFT
approach to reproduce molecular properties is not completely satisfying
\citep{sundararaman_computationally_2012,sundararaman_efficient_2014}.

Zhao \textit{et al} recently proposed the so-called Reaction Density
Functional Theory (RxDFT) \citep{tang_development_2018,cai_reaction_2019,tang_solvent_2020-1}
which is based on MDFT. In this approach the solute energy is computed
using eDFT. Then the solute is described classically by a set of Lennard-Jones
sites and point charges in a subsequent MDFT calculation to estimate
the solvation free energy. The classical charges are fitted to reproduce
the molecular electrostatic potential (ESP) computed in the QM calculation.
This is a first limitation since there is no unique way to determine
the ESP charges as illustrated by the variety of existing fitting
methods \citep{sigfridsson_comparison_1998,marenich_charge_2012,hu_fitting_2007}.
Another limitation is the absence of polarisation of the QM solute
under the influence of non homogeneous solvent density in RxDFT. The
ESP charges are fitted using electronic densities computed in vacuum
or using a CSM. This method thus actually consists in a MDFT calculation
with a force field where the charges have been reparametrised on a
prior eDFT calculation rather than in a self-consistent QM/MDFT procedure. 

The purpose of this paper is to address these limitations and to propose
a QM/MDFT procedure where the quantum part and the MDFT part are optimised
self-consistently. Moreover, in contrast to the common practise in
QM/MM approaches, the solute electronic density is directly used in
the MDFT calculation which prevent to use ill-defined atomic point
charges. The rest of the paper is organised as follow, first we present
the theoretical aspects of the coupling between QM and MDFT before
testing the proposed methodology on two commonly-used benchmarks.
We first focus on the solvation of an eDFT water molecule in MDFT
water before addressing an aqueous chemical reaction namely a symmetric
nucleophilic substitution between chloromethane and chloride.

\section{Theory\label{sec:Theory}}

\begin{figure}
\centering{}\includegraphics[width=0.4\paperwidth]{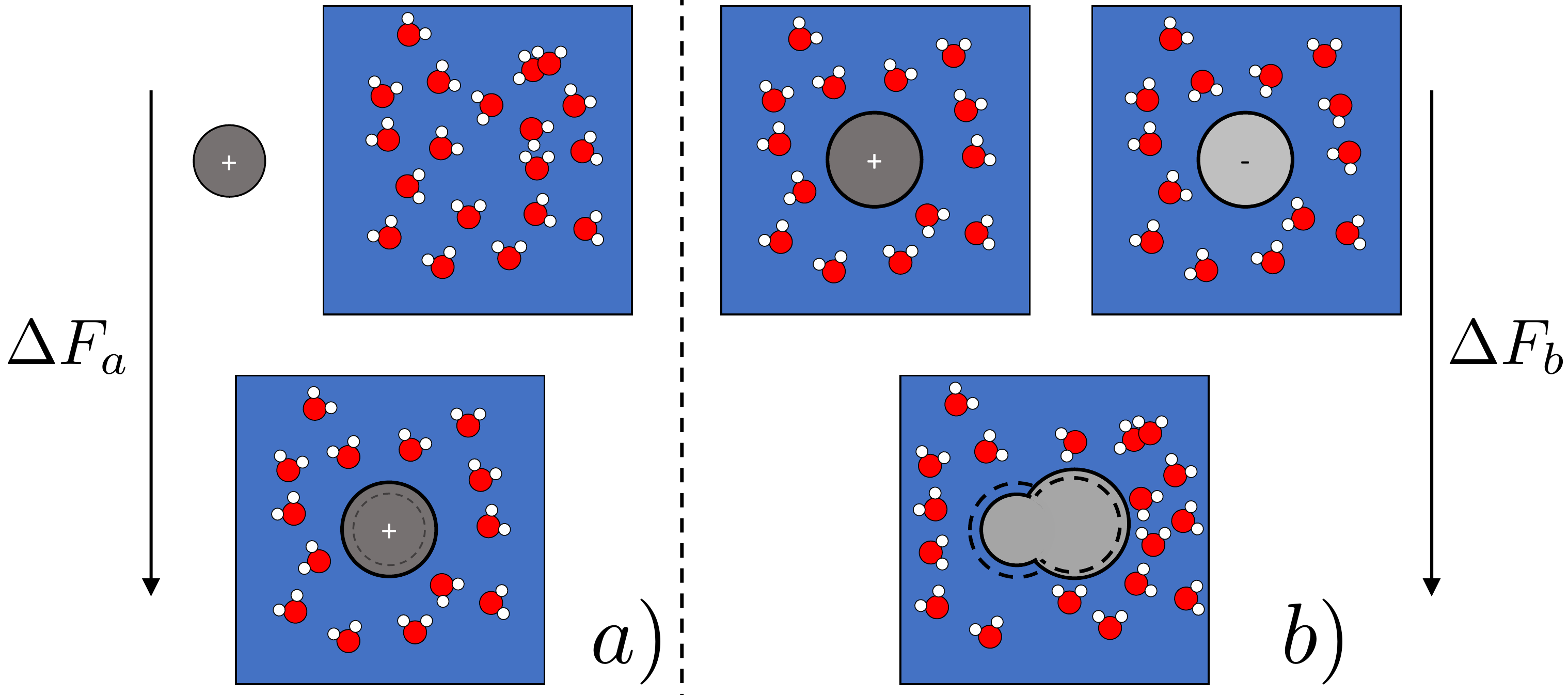}\caption{Thermodynamic scheme for the computation of solvation free energy
(left) and reaction free energy (right). The electronic density of
the solute is schematised by a solid black line. Dashed line represent
the electronic density in the previous state. \label{fig:shema-solvation-FE}}
\end{figure}

The solvation free energy (SFE) $\Delta F$ can be defined as the
difference between the free energy of the solute-solvent system and
the sum of the free energies of the solute in vacuum and of the pure
solvent as depicted in figure \ref{fig:shema-solvation-FE}.a. It
is a key quantity to understand chemistry in solution. For instance,
SFE difference between the products and the reactants, figure \ref{fig:shema-solvation-FE}.b,
is directly linked to the equilibrium constant of the reaction. Similarly,
the difference between the SFE of the same solute in two different
solvents allow to compute partition constant. The aim of this paper
is to propose a joint self-consistent eDFT/MDFT approach to evaluate
the SFE.

We first start by our standard formulation of molecular density functional
theory. In MDFT, the solvent molecules are assumed to be rigid entities
interacting through a classical force field. Since molecules are rigid,
the knowledge of the position of their centre of mass (COM) $\bm{r}$
and their orientation $\bm{\Omega}$ are sufficient to fully described
molecule coordinates. In this paper we only consider SPC/E water as
a solvent but we proposed functional for other molecular fluids such
as acetonitrile in the past \citep{borgis_molecular_2012}. The DFT
ansatz states that for any external perturbation it is possible to
write a unique functional $F$ of the solvent density $\rho$ \citep{evans_nature_1979}.
At its minimum, which is reached for the equilibrium density $\rho_{\text{eq}}$,
the functional $F$ equals the SFE $\Delta F$. MDFT is thus particularly
appropriate to compute SFE since it requires a functional minimisation
while brute force MD would require a costly sampling. The functional
$F$ is usually written as 
\begin{equation}
F[\rho]=F_{\text{id}}[\rho]+F_{\text{ext}}[\rho]+F_{\text{exc}}[\rho].\label{eq:F=00003DFid+Fexc+Fext-1}
\end{equation}
 The solvent density $\rho(\bm{r},\bm{\Omega})$ is a 6D field that
depends on the space coordinate and the orientation $\bm{\Omega}$.

In equation \ref{eq:F=00003DFid+Fexc+Fext-1} the ideal term $F_{\text{id}}$
corresponds to the entropic term of the non-interacting fluid \citep{hansen_theory_2006}.
The third term $F_{\text{exc}}$ is due to solvent-solvent interaction
\citep{jeanmairet_molecular_2013-1}. In this work we use the expression
proposed by Ding et al for SPC/E \citep{ding_efficient_2017}, which
corresponds to the so-called hypernetted chain for the excess functional
$F_{\text{exc}}$. The remaining term $F_{\text{ext}}$ is due to
the external perturbation acting on the liquid, here the solute. This
last term represents the solute-solvent interaction and can formally
be written as\foreignlanguage{english}{
\begin{equation}
F_{\text{ext}}[\rho]=\iint\rho\left(\bm{r},\bm{\bm{\Omega}}\right)V_{\text{ext}}(\bm{r},\bm{\Omega})d\bm{r}d\bm{\Omega}.\label{eq:Fext-1}
\end{equation}
}where $V_{\text{ext}}$ is the external energy density.

In our previous work \citep{jeanmairet_molecular_2013,jeanmairet_molecular_2013-1,jeanmairet_molecular_2015,jeanmairet_molecular_2016,jeanmairet_molecular_2019,jeanmairet_study_2019,levesque_solvation_2012}
the solute was described by a classical force field, usually a set
of Lennard-Jones and point charges. Here we describe the solute quantum
mechanically using eDFT. Using the DFT ansatz \citep{hohenberg_inhomogeneous_1964},
it exists a functional $F_{e}$ of the electronic density $\rho_{e}$
which is equal to the ground state energy at its minimum. The electrostatic
interaction between the quantum solute and the classical solvent can
easily be expressed in a mean field way
\begin{equation}
E_{\text{ES}}[\rho,\rho_{e}]=\iint\frac{\sigma_{\text{V}}(\bm{r})\sigma_{\text{U}}(\bm{r}^{\prime})}{4\pi\epsilon_{0}\left|\bm{r}-\bm{r}^{\prime}\right|}d\bm{r}d\bm{r}^{\prime}\label{eq:F_ES}
\end{equation}

where $\sigma_{\text{V}}$ is the charge density of the solvent and
$\sigma_{\text{U}}$ is the charge density of the solute. Each of
these charge density is related to the corresponding particle density.
The solute charge density is
\begin{equation}
\sigma_{\text{U}}(\bm{r})=\sum_{i}Z_{i}\delta(\bm{r}-\bm{r}_{i})-\rho_{e}(\bm{r})\label{eq:sigma_u}
\end{equation}
where $Z_{i}$ is the atomic number of nucleus $i$ located in $\bm{r}_{i}$.
The solvent charge density can be expressed as
\begin{equation}
\sigma_{\text{V}}(\bm{r})=\iint\rho(\bm{r},\bm{\Omega})\sigma(\bm{r}-\bm{r}^{\prime},\bm{\Omega})d\bm{r}^{\prime}d\bm{\Omega}.\label{eq:sigma_v}
\end{equation}
where $\sigma(\bm{r},\bm{\Omega})$ is the charge density of a water
molecule taken at the origin with orientation $\bm{\Omega}$.

The electrostatic contribution to the external term of equation \ref{eq:Fext-1}
can thus be computed injecting equations \ref{eq:sigma_u}-\ref{eq:sigma_v}
in equation \ref{eq:F_ES}. However, short-range repulsion and dispersion
interactions are not taken into account. To do so, similarly to QM/MM
calculations, we resort to Lennard-Jones sites located on nuclei of
the solute. Since there is no prescriptions on how to choose the Lennard-Jones
parameters the common practice is to resort to generic force fields
such as OPLS \citep{tirado-rives_qm/mm_2019} or CHARMM \citep{riccardi_importance_2004}.
However, the solvation free energy and the solvation structure depend
on the LJ parameters \citep{tu_effect_1999,riccardi_importance_2004}.
That is why, as any QM/MM calculation, the present approach cannot
be considered as being truly ab-initio. A more elegant and ab-initio
way would be to use some electron-solvent pseudo-potential to account
for the repulsion-dispersion interactions \citep{vaidehi_quantummechanical_1992}.
This strategy has been widely applied to study solvated electrons
in liquids or clusters \citep{schnitker_electronwater_1987,turi_analytical_2002,mones_new_2010}.
Eventually, the external term of the functional can be written as
\begin{equation}
F_{\text{ext}}[\rho,\rho_{e}]=E_{\text{ES}}[\rho,\rho_{e}]+\iint\rho\left(\bm{r},\bm{\bm{\Omega}}\right)V_{\text{LJ}}(\bm{r},\bm{\Omega})d\bm{r}d\bm{\Omega}.\label{eq:FextQM/MM}
\end{equation}
with
\begin{eqnarray*}
V_{\text{LJ}}(\bm{r},\bm{\Omega}) & = & \sum_{i\in\text{solute}}\sum_{j\in\text{solvent}}4\epsilon_{ij}\left[\left(\frac{\sigma_{ij}}{\left|\bm{r}+\bm{r}_{j\Omega}-\bm{r}_{i}\right|}\right)^{12}\right.\\
 &  & -\left.\left(\frac{\sigma_{ij}}{\left|\bm{r}+\bm{r}_{j\Omega}-\bm{r}_{i}\right|}\right)^{6}\right]
\end{eqnarray*}
where $\epsilon_{ij}$ and $\sigma_{ij}$ are the mixed Lennard-Jones
parameters using the Lorentz-Berthelot rules, $\bm{r}_{i}$ is the
position of the $i^{\text{th}}$ site of the solute and $\bm{r}_{j\Omega}$
denotes the position with respect to the COM of site $j$ of a solvent
molecule located in $\bm{r}$ with orientation $\bm{\Omega}$.

As opposed to our previous work where the solute was described classically,
the free energy of the solute is modified when transferred from the
gas phase to the solute. We approximate this free energy difference
$\Delta F_{\text{QM}}$ by the energy difference at $T=0\ K$ which
is much easier to compute. This neglects the nuclear and electronic
fluctuations.
\begin{equation}
\Delta F_{\text{QM}}[\rho_{e}]\approx\Delta E_{\text{QM}}[\rho_{e}]=F_{e}[\rho_{e}]-F_{e}[\rho_{e}^{\text{vac}}]\label{eq:deltaE_qm_old-1}
\end{equation}
where $\rho_{e}^{\text{vac}}$ is the equilibrium electronic density
in vacuum. Finally, using equations \ref{eq:F=00003DFid+Fexc+Fext-1},\ref{eq:FextQM/MM}
and \ref{eq:deltaE_qm_old-1} the solvation free energy can be computed
by minimising the functional 
\begin{equation}
F\left[\rho_{e},\rho\right]=\Delta E_{\text{QM}}[\rho_{e}]+F_{\text{id}}[\rho]+F_{\text{ext}}[\rho]+F_{\text{exc}}[\rho_{e},\rho]\label{eq:F_the_one}
\end{equation}
 with respect to the electronic density $\rho_{e}(\bm{r})$ and the
solvent density $\rho(\bm{r},\bm{\Omega})$.

Instead of carrying the joint minimisation we adopt a simpler strategy.
First, the electronic functional is minimised in vacuum. The equilibrium
electronic density is then used in the MDFT calculation to compute
the electrostatic contribution to external term using equation \ref{eq:F_ES}.
After minimisation of the MDFT functional, the equilibrium solvent
charge density is used to compute the electrostatic external potential
acting on the electronic density using equation \ref{eq:F_ES}. The
electronic functional is minimised and a new electronic density is
obtained. This process is repeated until both the electronic energy
of equation \ref{eq:deltaE_qm_old-1} and the solvation free energy
of equation \ref{eq:F=00003DFid+Fexc+Fext-1} are converged to a given
threshold. Using this procedure, the electrostatic energy of equation
\ref{eq:F_ES} is computed twice, once in the electronic DFT calculation
and once in the MDFT one. These two values can be compared as a sanity
check to verify convergence.

We insist on the fact that the full electronic density of the quantum
solute is used in the computation of electrostatic interaction between
the QM and the MM part in equation \ref{eq:F_ES}. It differs from
the strategy usually adopted in QM/MM calculation that consists in
computing partial point charges from the electronic density \citep{lin_qm/mm:_2007}.
Since there is no unique way to determine these charges \citep{marenich_charge_2012,sigfridsson_comparison_1998,sifain_discovering_2018}
and since it is difficult to evaluate their quality a posteriori it
is advantageous to circumvent this parametrisation and work directly
with the electronic density.

The self-consistent optimisation of electronic density $\rho_{e}(\bm{r})$
and the solvent density $\rho(\bm{r},\bm{\Omega})$ when minimising
equation \ref{eq:F_the_one} allows to account for the mutual polarisation
of the solute and the solvent environment. This is a clear improvement
with respect to methods that consists in single QM calculation in
vacuum followed by a single liquid state theory calculation such as
RxDFT \citep{tang_development_2018,cai_reaction_2019,tang_solvent_2020-1}.
Indeed, such methods neglect the polarisation of the solute by the
solvent density which is properly accounted with the present approach.

A last strength of the present approach is that it retains a proper
description of the solvent at the molecular level. The equilibrium
solvent density contains a detailed 3D picture of the solvent location
and orientations around the solute which is not accessible to continuum
models or simpler liquid state theories.

\section{Applications\label{sec:Applications}}

\subsection{Water in water.}

As a first test of the QM/MDFT framework introduced in \ref{sec:Theory},
we focus on the ``water in water'' system. A single water molecule,
referred as the solute, is immersed in water solvent described by
MDFT. The solute is treated at the eDFT level with the PBE functional.
Calculations are run using the GPAW package \citep{Enkovaara_2010,ase-paper,ISI:000226735900040}.
Wave functions are expanded on a real space grid. The volume of the
simulation cell is $5\times5\times5$~$\textrm{Å}^{3}$ and details
about the grid resolution are given after. The geometry of the solute
is the one of the SPC/E molecule.

Solvent calculations are done using our homemade MDFT code, the water
model is SPC/E. We used a cubic cell of $25\times25\times25\ \textrm{Å}^{3}$.
The orientational space $\text{SO}(3)$ is discretised with 196 orientations,
the space grid resolution is specified further. Our homemade fortran
written MDFT program was interfaced with Python using f90wrap which
makes the coupling with GPAW easy. 

Dispersion forces are modelled using Lennard-Jones sites located on
the solute atoms in the MDFT calculation. The Lennard-Jones parameters
of the oxygen of the solute are the same as in SPC/E. The Lennard-Jones
parameters on hydrogens are $\sigma=1.0\ \textrm{Å}$ and $\epsilon=0.0234$
kJ.mol$^{-1}$. This almost non attractive Lennard-Jone site prevent
numerical divergence due to ``unshielded'' charges, this trick has
already been used in RISM-SCF studies \citep{ten-no_hybrid_1993,tenno_reference_1994}.

We first check the validity of the implementation by comparing the
external electrostatic energies obtained in the QM and in the MDFT
calculations according to equation \ref{eq:F_ES}. For this test there
are $n^{3}$ points in the QM grid and $8\ n^{3}$ in the MDFT one
where $n=32,40,48,64$. The convergence criterion on the relative
variation is 10$^{-4}$ for both the QM energy and the MDFT free energy.
Results are reported in table \ref{tab:check-external-electrostatic}.
In all cases the electrostatic energies agree within 1\%. For $n=32$
the resolution of the grids are clearly not sufficient since the results
differs by 5 $\text{kJ.mol}^{-1}$ from the one obtained with finer
grids. If $n=32$ is omitted, the finer the grid the better is the
agreement between the two ways to evaluate the electrostatic energy.

\begin{table}
\centering{}%
\begin{tabular}{|c|c|c|c|}
\hline 
$n$ & QM $\left(\text{kJ.mol}{}^{-1}\right)$  & MDFT $\left(\text{kJ.mol}{}^{-1}\right)$  & $\frac{\text{QM}}{\text{MDFT}}$\tabularnewline
\hline 
\hline 
32 & -54.5077 & -54.4329 & 1.0014\tabularnewline
\hline 
40 & -59.5362 & -59.8865 & 0.9942\tabularnewline
\hline 
48 & -59.1667 & -59.3552 & 0.9968\tabularnewline
\hline 
64 & -59.1093 & -59.2150 & 0.9982\tabularnewline
\hline 
\end{tabular}\caption{Final external electrostatic energy of water in water computed according
to equation \ref{eq:F_ES}. The second column is the result obtained
with the QM code, the result obtained with MDFT is displayed in the
third column. The fourth column show the ratio of the second and third
column. There are $n^{3}$ nodes on the QM grid and $8n^{3}$ on the
MDFT grid. \label{tab:check-external-electrostatic}}
\end{table}

\begin{figure}
\centering{}%
\begin{tabular}{c}
\includegraphics[width=0.4\paperwidth]{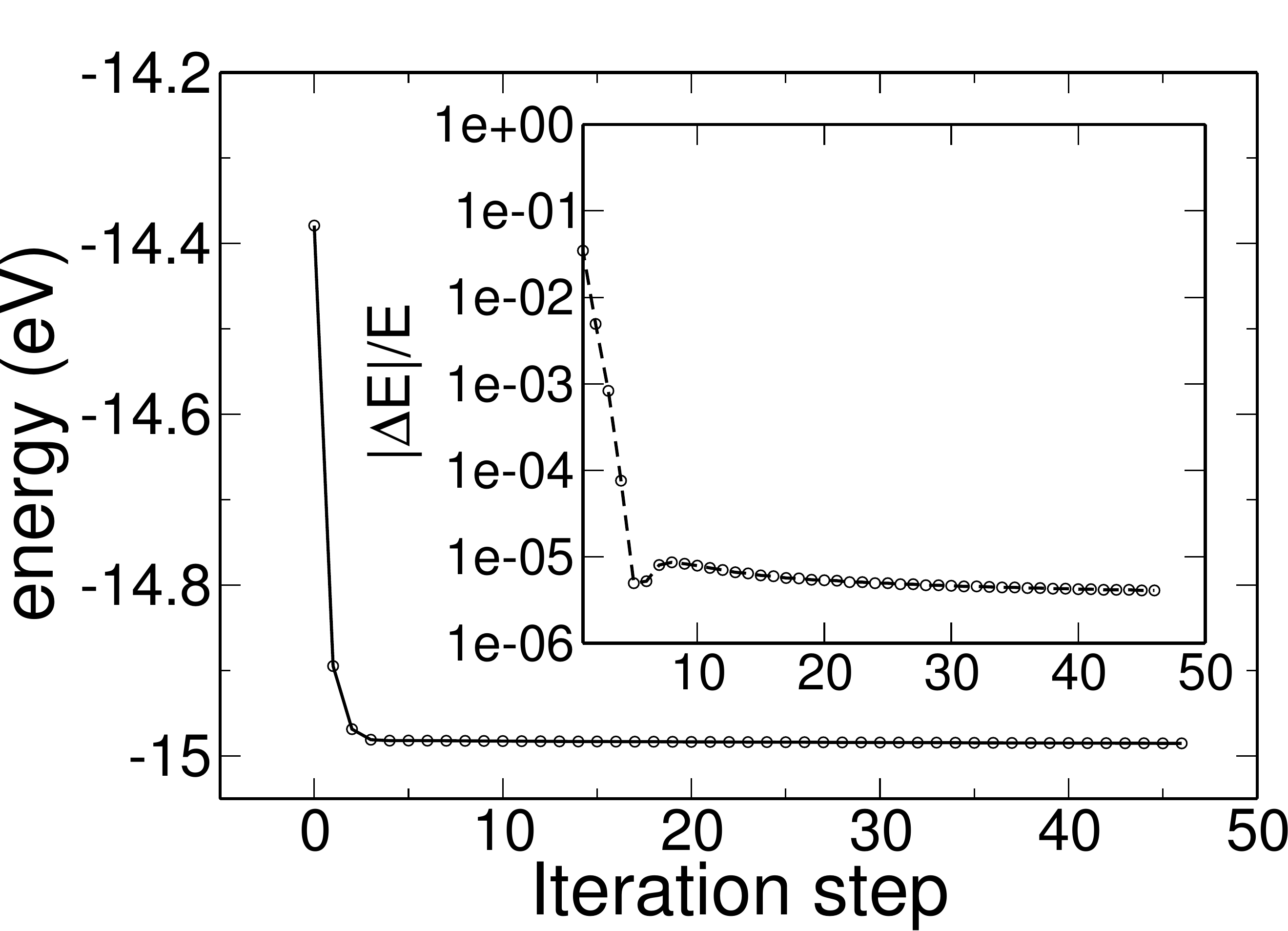}\tabularnewline
\includegraphics[width=0.4\paperwidth]{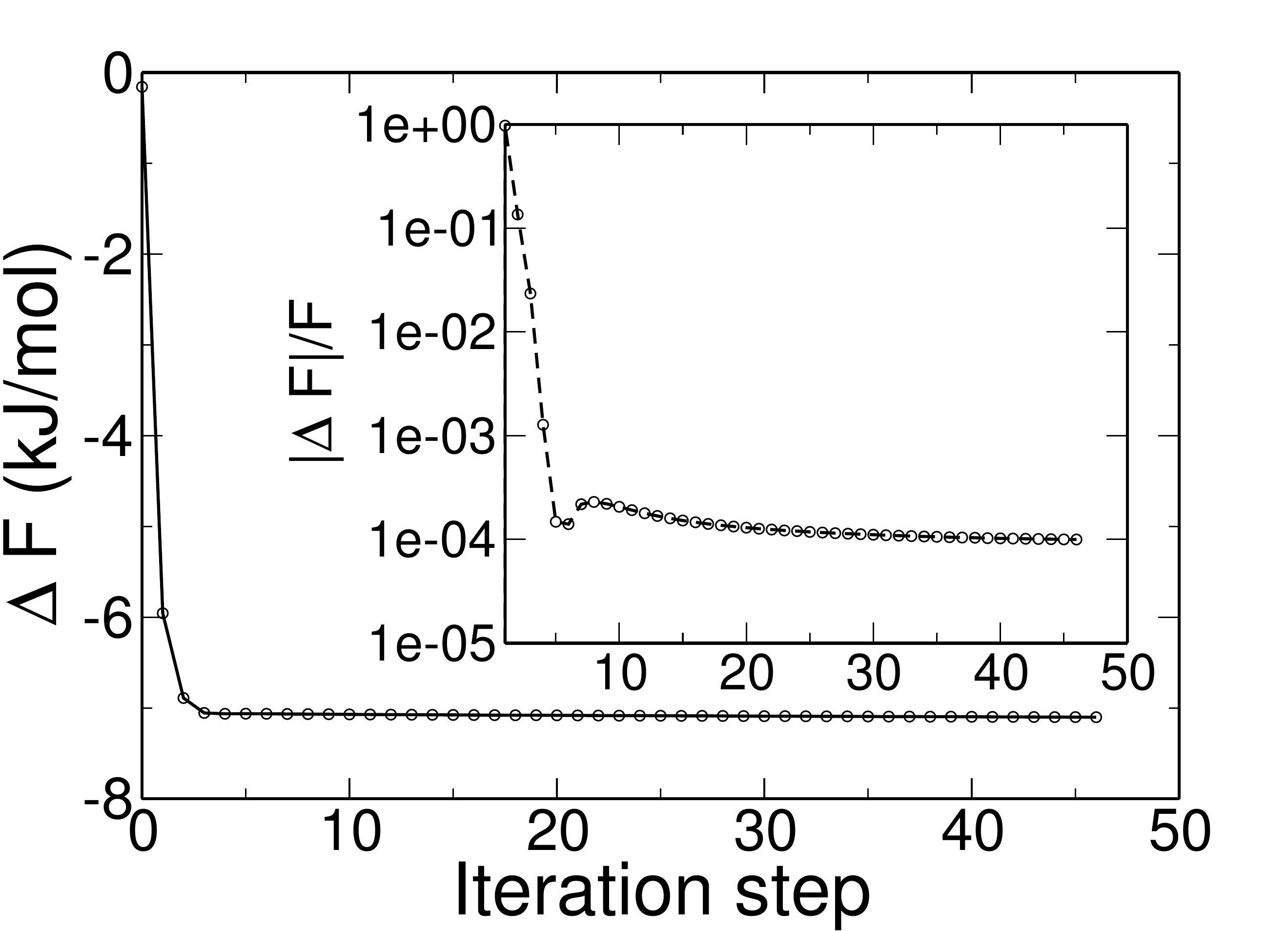}\tabularnewline
\end{tabular}\caption{Energy of the quantum solute as computed by eDFT (top) and its solvation
free energy as computed by MDFT (down) as a function of the iteration
step. In insets are the relative differences of these quantities.\label{fig:Energy-convergence-water}}
\end{figure}

After this numerical test we run calculation on the same system with
$48^{3}$ grid nodes on the QM grid and $120^{3}$ nodes on the MDFT
space grid. We use a convergence criterion of 10$^{-4}$ on the relative
variation of QM energy and MDFT free energy. The electronic energies
and solvation free energies as a function of the iteration step are
displayed in figure \ref{fig:Energy-convergence-water}. While it
requires 42 iterations to reach the required criterion only 5 iterations
are necessary to reach a criterion of $10^{-3}$. As expected, solvation
stabilises the solute: the QM energy at convergence is $0.58\ \text{eV}$
lower than in the initial state \textit{i.e.} in vacuum. In a similar
way, the solvation free energy computed by MDFT is reduced by $6.6\ \text{kJ.mo\ensuremath{\text{l}^{-1}}}$
when the electronic density is optimised. 

Moreover, the polarisation of the solvent influences the electronic
density of the solute. This effect can only be captured if the electronic
density and the solvent density are optimised self-consistently.

The solvation free energy of water computed using equation \ref{eq:F_the_one}
is $-3.4\ \text{kJ.mo\ensuremath{\text{l}^{-1}}}$. It is clearly
overestimated when compared to the experimental value of $-26.5\ \text{kJ.mol}{}^{-1}$
but also to the value of $-29.5\ \text{kJ.mol}{}^{-1}$ which is the
one of the SPC/E model computed using MD. Using the continuum solvent
model implemented in GPAW \citep{held_simplified_2014} gives a solvation
free energy of $-27.0\ \text{kJ.mol}{}^{-1}$ in better agreement
with the experimental value. The overestimation of solvation free
energy is a known problem of the HNC functional as already illustrated
for the TIP3P model \citep{luukkonen_hydration_2020} and is mostly
due to the quadratic form of the functional, which cause a tremendous
overestimation of the pressure \citep{jeanmairet_molecular_2013,jeanmairet_molecular_2015}.
We have proposed several physically based bridge functionals to correct
this flaw and go beyond the HNC level. It is also possible to stay
at the HNC level and simply use a one parameter a posteriori cavity
correction of the SFE \citep{luukkonen_hydration_2020}. Using this
simple correction, we obtain a solvation free energy of $-20.2\ \text{kJ.mol}{}^{-1}$,
in better agreement with the experimental value but still overestimated. 

Unfortunately, the imperfection of the functional is not the sole
defect of our calculation. Predictions of the solvation free energy
are also quite sensitive to the the choice of Lennard-Jones parameters.
To illustrate this point we computed the solvation free energy of
water changing only the Lennard-Jones site on the Oxygen of the solute.
We have taken the values of $\sigma$ and $\epsilon$ for some popular
water model: SPC/E, OPC, TIP3P and TIP4P. We emphasise that the geometry
of the solute is not changed. The SFE computed using these parameters
are displayed in table \ref{tab:free_ener_Water_LJ} and they vary
by up to 1.5 $\text{kJ.mol}^{-1}$. 
\begin{table}
\centering{}%
\begin{tabular}{|c|c|c|c|}
\hline 
 & $\sigma$ $\left(\textrm{Å}\right)$  & $\epsilon$ $\left(\text{kJ.mol}{}^{-1}\right)$  & $\Delta F$ $\left(\text{kJ.mol}{}^{-1}\right)$ \tabularnewline
\hline 
\hline 
TIP3P & 3.15061 & 0.6364 & -20.3\tabularnewline
\hline 
TIP4P & 3.1589 & 0.7749 & -19.5\tabularnewline
\hline 
OPC3 & 3.165 & 0.9945 & -18.8\tabularnewline
\hline 
SPC/E & 3.17427 & 0.65 & -20.2\tabularnewline
\hline 
\end{tabular}\caption{Solvation Free energy of water obtained using different Lennard-Jones
parameters for the solute described using eDFT, its geometry is always
the one of SPC/E. \label{tab:free_ener_Water_LJ}}
\end{table}

\begin{figure}
\centering{}%
\begin{tabular}{c}
\includegraphics[width=0.4\paperwidth]{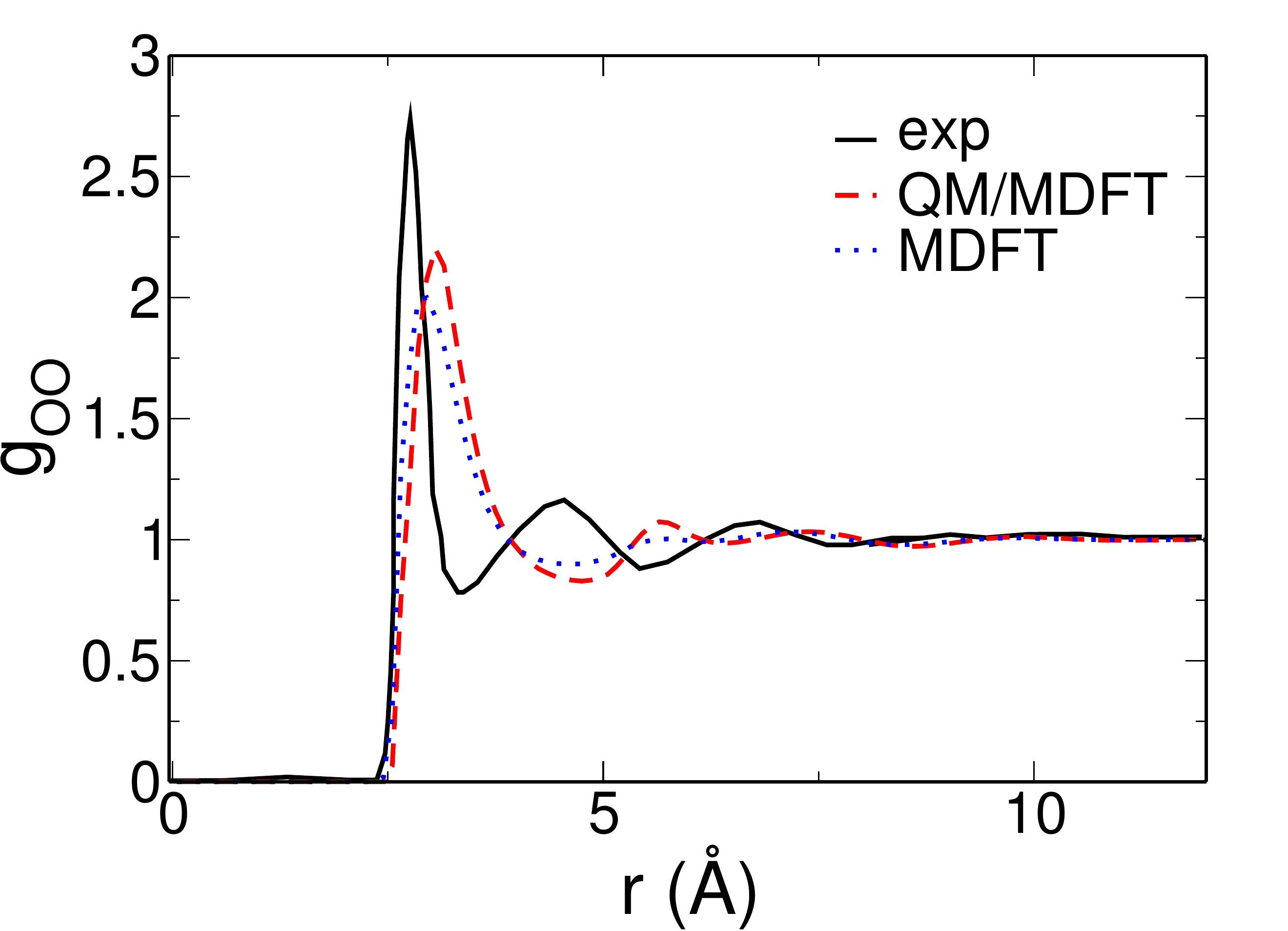}\tabularnewline
\includegraphics[width=0.4\paperwidth]{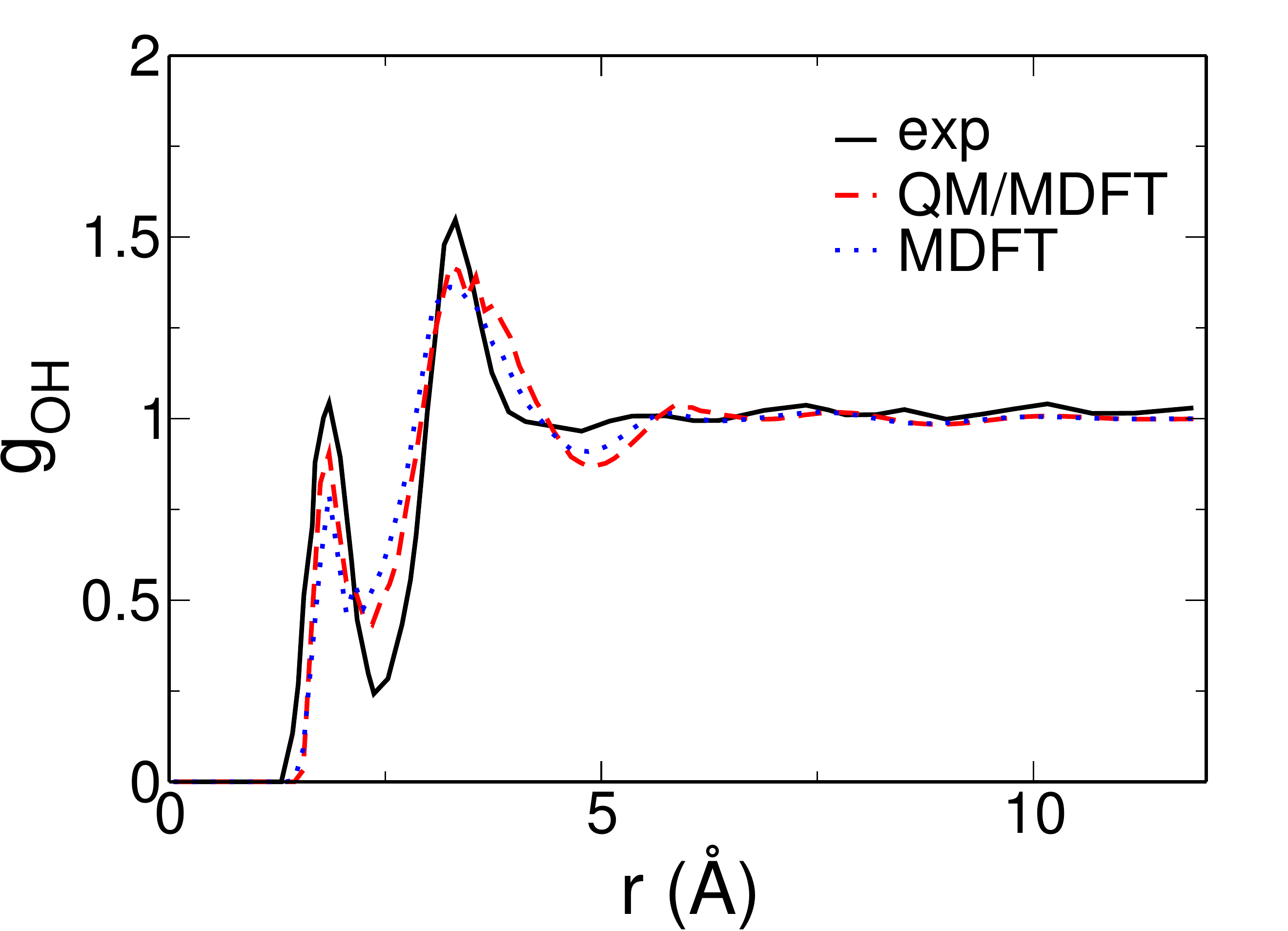}\tabularnewline
\end{tabular}\caption{Radial distributions functions between the O site of the solvent and
the O (top) or H (down) site of the solute. Results of the HNC functional
for the classical SPC/E solute are in dotted blue while the one obtained
with a QM solute are in dashed red. For comparison we reported the
experimental results by Soper in full black \citep{soper_radial_2000}.
\label{fig:rdf-water}}
\end{figure}

After examining the energetics, we now turn to the solvation structure.
The radial distribution functions (rdf) between oxygen of the solvent
and atomic sites of the solute are displayed in figure \ref{fig:rdf-water}.
First, we recall that the agreement between the experimental rdf and
the one computed by MD for the SPC/E model is good \citep{soper_radial_2000,mark_structure_2001}.
The agreement is less satisfying for the rdf predicted by MDFT using
the HNC functional as previously reported for SPC/E and TIP3P \citep{jeanmairet_molecular_2013-1,luukkonen_hydration_2020}.
Indeed, the first peak of the OO rdf is overestimated and slightly
shifted toward the long distances while the second and third peaks
are underestimated and markedly shifted towards the long distances.
The agreement is much better for the OH rdf since the two first peaks
are found at the right places even if there are underestimated as
is the depletion between the two peaks. Since the present approach
uses the HNC functional with no modifications, the same defects are
found on the rdf computed using the QM/MDFT framework. Using a quantum
solute tend to improves the intensity of the peaks: the two first
peaks of the OH and OO rdf increases. However the peaks of the OH
rdf are still underestimated and the depletion between the two first
peaks of the OH rdf remains too small. Considering the position of
the peaks, there is no improvement on the OH rdf and it even worsen
the OO rdf where the first maximum is shifted further towards the
large distances.

Overall the radial distribution functions computed using the QM/MDFT
approach remain similar to the one obtained using MDFT on a classical
SPC/E solute. We can expect that bridge functionals improving the
structural properties on classical systems to be transferable to QM/MDFT
calculations.

While rdf function is a practical way to examine solvation structure
it only contains spherically averaged information. This is not the
case of the 3D densities that can be computed with MDFT. We estimate
the solvent charge $\sigma_{\text{CSM}}$ using the CSM model implemented
in GPAW \citep{held_simplified_2014} and compare it to the 3D solvent
charge density $\sigma_{\text{V}}$ given by equation \ref{eq:sigma_v}.
We assume that the whole difference between the electrostatic potential
in CSM, $\Phi_{\text{CSM}}^{\text{ES}}$ and the one in vacuum $\Phi_{\text{vac}}^{\text{ES}}$
is due to the electrostatic potential generated by solvent $\Phi_{\text{V}}^{\text{ES}}$
\begin{equation}
\Phi_{\text{V}}^{\text{ES}}=\Phi_{\text{CSM}}^{\text{ES}}-\Phi_{\text{vac}}^{\text{ES}}.
\end{equation}
This electrostatic potential $\Phi_{\text{V}}^{\text{ES}}$ is linked
to the solvent charge $\sigma_{\text{CSM}}$ through a Poisson equation
\begin{equation}
\Delta\Phi_{\text{V}}^{\text{ES}}=-\frac{\sigma_{\text{CSM}}}{\epsilon_{0}}\label{eq:poisson_bound_charge}
\end{equation}
where $\epsilon_{0}$ is the vacuum permitivitty.

Of course the charge density $\sigma_{\text{CSM}}$ actually does
not solely contain the contribution due to the dielectric response
of the solvent. The modification of the electronic density also impacts
the electrostatic potential. Moreover, in equation \ref{eq:poisson_bound_charge}
we simply used the vacuum permitivitty while we should have used the
spatially varying permittivity entering the continuum model. Thus,
the charge density $\sigma_{\text{CSM}}$ is simply a qualitative
tool to visualise solvation effect in the CSM calculation. 

\begin{figure}
\centering{}\includegraphics[width=0.4\paperwidth]{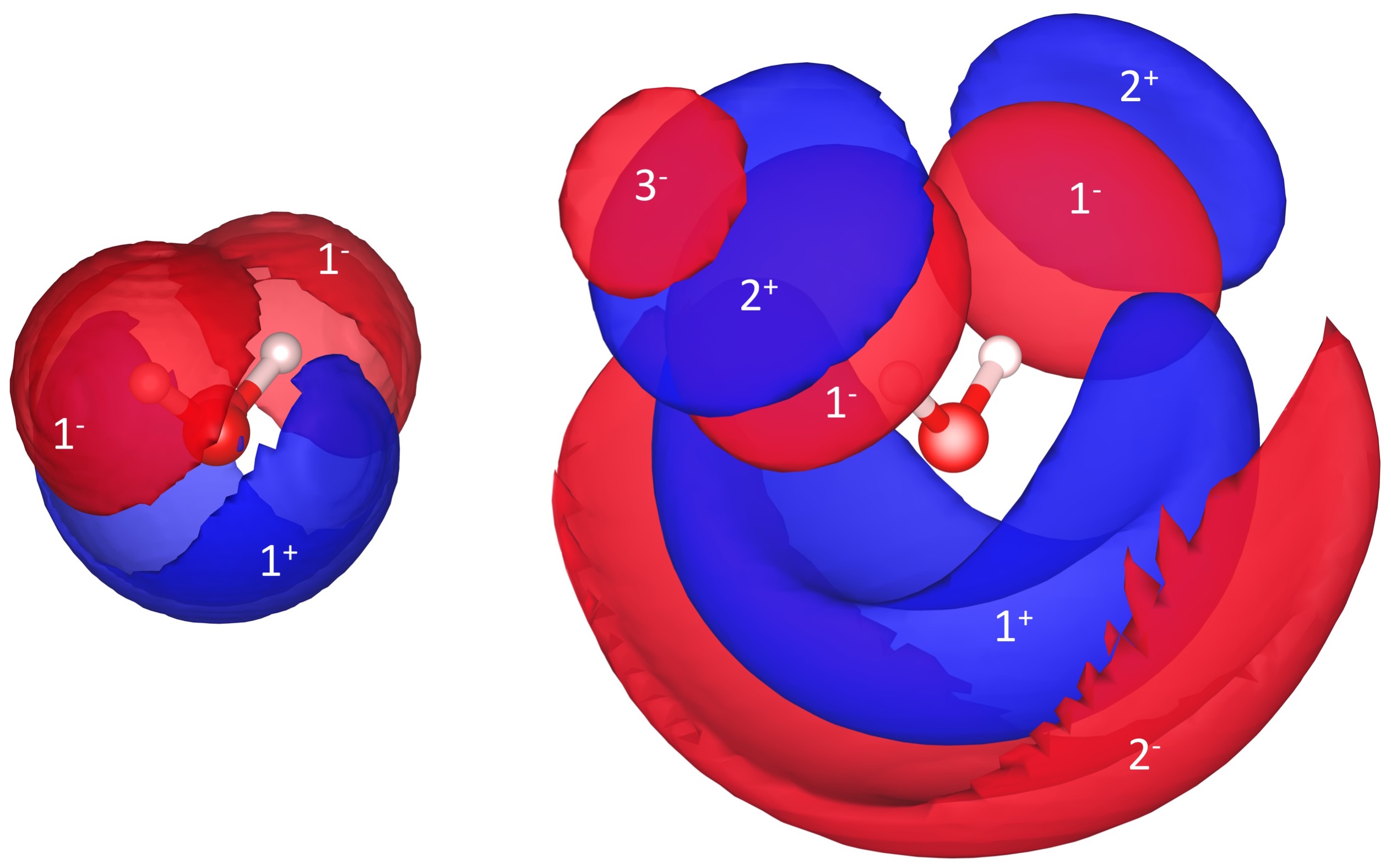}\caption{Isosurfaces of solvent charge densities: positive surfaces are displayed
in blue and negative ones in red. The left figure have been obtained
with the CSM calculation using equation \ref{eq:poisson_bound_charge}.
The right one have been obtained with MDFT with equation \ref{eq:sigma_v}.
\label{fig:Isosurfaces-of-solvent}}
\end{figure}

In figure \ref{fig:Isosurfaces-of-solvent} we compare the charge
densities obtained with CSM and with MDFT. The solvent charge density
$\sigma_{\text{V}}$ predicted by MDFT is more structured than $\sigma_{\text{CSM}}$:
there are additional lobes. To ease the discussion, we identify the
lobes in figure \ref{fig:Isosurfaces-of-solvent} by their distance
with respect to the centre of mass of the solute. The closest positive
lobe is denoted $1^{+}$, the second negative $2^{-}$, etc. 

The lobes $1^{+}$ and $1^{-}$ are similar in the CSM calculation
and in the MDFT calculation. However, the charge distribution obtained
by CSM is located on the cavity surface while the lobes obtained with
MDFT have 3D shapes.

$1^{+}$ and $1^{-}$ correspond to the first solvation shell and
not surprisingly we observe that the solvent is polarised such as
positives charges appear close to the oxygen atom while negative charges
develop close to the hydrogen atoms. In the MDFT results, a second
set of lobes exist. They have a shape similar to the one of the first
lobes but with opposite charge and they are located in their vicinity,
e.g. $2^{+}$ is close to $1^{-}$. We estimated the distance between
$1^{+}$ and $2^{-}$ and between $2^{+}$ and $1^{-}$ by measuring
distances between several pairs of points pertaining to each isosurface.
We found that both pairs of isosurfaces are roughly distant by $1\ \textrm{Å}$.
This is of the order of the OH bond length in the SPC/E water model,
thus each pairs of oppositely charge isosurfaces belong to the first
solvation shell of water. We recover the tetrahedral order with preferential
orientation of water around the water solute molecule. 

While the dipole moment of the water molecule is well known to be
1.85 D in the gas phase \citep{clough_dipole_1973} its value in the
liquid have been more controversial with simulation prediction ranging
from 2.1 to 3.1 D \citep{gubskaya_total_2002}. In the beginning of
the 2000's several independent experimental studies have found a value
of 2.9-3.0 D for the dipole moment in the liquid \citep{badyal_electron_2000,gubskaya_total_2002}.
In table \ref{tab:dipole-water} we display the dipole moment of a
water molecule in gas and liquid phase. We obtain a dipole of 1.9
D with the PBE functional, in good agreement with the experimental
value. The value in solution is estimated to 2.3 D with the continuum
solvation model. This is clearly underestimated with respect to the
experimental value but this falls within the range of values predicted
using simulations and QM/MM approaches \citep{rocha_efficient_2001}.
Note that this value is interestingly close to 2.35 D which is the
one of the SPC/E model \citep{mark_structure_2001}. With the MDFT
approach we do obtain an enhancement of the dipole of the water molecule
in solution. After a single MDFT calculation most of the polarisation
is recovered with a dipole of 2.2 D. After converging the self consistent
cycle, the dipole increases to 2.3 D. Once again, this shows the importance
of the self consistent optimisation of electronic and solvent densities
to account for their mutual polarisation.

Note that the same value of the dipole is obtained using any of the
Lennard-Jones parameters of table \ref{tab:free_ener_Water_LJ}. It
should be noted that several AIMD simulations of liquid water using
the PBE functional have found a dipole value of 2.9-3.2 D \citep{allesch_first_2004,mcgrath_spatial_2007}
which is in good agreement with experiments. The underestimation of
the dipole moment in the liquid phase in the MDFT calculation may
have several origin. First, it has been demonstrated using ab initio
MD that the dipole moment of rigid water molecules tends to be smaller
than when the geometry is allowed to relax \citep{allesch_first_2004}.
Second, as opposed to AIMD there is no electronic polarisation of
the solvent molecules since we use a non-polarisable model. A third
reason might be the limitation of the HNC functional. As illustrated
in figure \ref{fig:rdf-water}, the first peak of the oxygen rdf is
broaden, reduced and shifted further from the solute with the HNC
functional with respect to experiments while the hydrogen rdf more
or less agree. Consequently, the charge distribution of the solvent
predicted by the HNC functional is smoother. As a consequence, the
electric field generated by the solvent is reduced and the water solute
is less polarised than in experiment. 

\begin{table}
\centering{}%
\begin{tabular}{|c|c|c|}
\hline 
$\mu$(D) &  vacuum & solution\tabularnewline
\hline 
\hline 
exp. & 1.85 & 2.9-3.0\tabularnewline
\hline 
CSM & 1.9 & 2.3\tabularnewline
\hline 
MDFT & 1.9 & 2.3 (2.2)\tabularnewline
\hline 
\end{tabular}\caption{Molecular dipole of the water molecule in the gas phase and in the
liquid. The vacuum value of 1.9 D correspond to a unique calculation.
The value obtained with a QM calculation followed by a unique MDFT
calculation is displayed in parenthesis. \label{tab:dipole-water}}
\end{table}
As a conclusion, this study of water in water with the QM/MDFT approach
is encouraging. We are able to recover the tetrahedral structure of
the first solvation shell around the solute and the enhancement of
the water dipole in the liquid phase. However, some defects of the
HNC functional are still present. In particular the solvation shell
is not structured enough. There is still room to improve the functional,
a natural route being to introduce an appropriate bridge term. These
defects should not obliviate the potential of the method due to its
computational efficiency with respect to AIMD. It took 25 minutes
on a 32 CPU desktop machine to carry the full self-consistent QM/MDFT
cycle, computing the chemical potential of water using AIMD would
require to use enhance sampling methods and take tens of thousands
of CPU-hours. To further illustrate the interest of the approach,
we now turn our attention to the prototypical symmetric $\text{S}{}_{\text{N}}2$
reaction between chloromethane and a chloride anion.

\subsection{$\text{S}{}_{\text{N}}2$ reaction}

Due to its extensive use as a tool to switch functional group in organic
molecules, many experimental and theoretical studies have been dedicated
to study the $\text{S}{}_{\text{N}}2$ reaction. Because the nucleophile
is very often an anion, solvent effect can deeply modify the free
energy profile of the reaction. Thus, a smart choice of solvent can
modulate the reactivity and the selectivity of the reaction \citep{hughes_55._1935,gertner_nonequilibrium_1989,ensing_solvation_2001}.
This explains that many studies have been dedicated to investigate
solvation effect in $\text{S}{}_{\text{N}}2$ reaction by simulation
\citep{bento_e2_2008,chandrasekhar_sn2_1984,chandrasekhar_theoretical_1985,gao_two-dimensional_1993,ensing_solvation_2001,kuechler_quantum_2014,gertner_nonequilibrium_1989,tirado-rives_qm/mm_2019,zhao_density_2010}.
In particular, QM/MM is a method of choice since it allows to account
for bond breaking/formation between the nucleophile and the electrophile
while keeping a realistic description of the solvent with a tractable
numerical cost.

\begin{figure}
\centering{}\includegraphics[width=0.4\paperwidth]{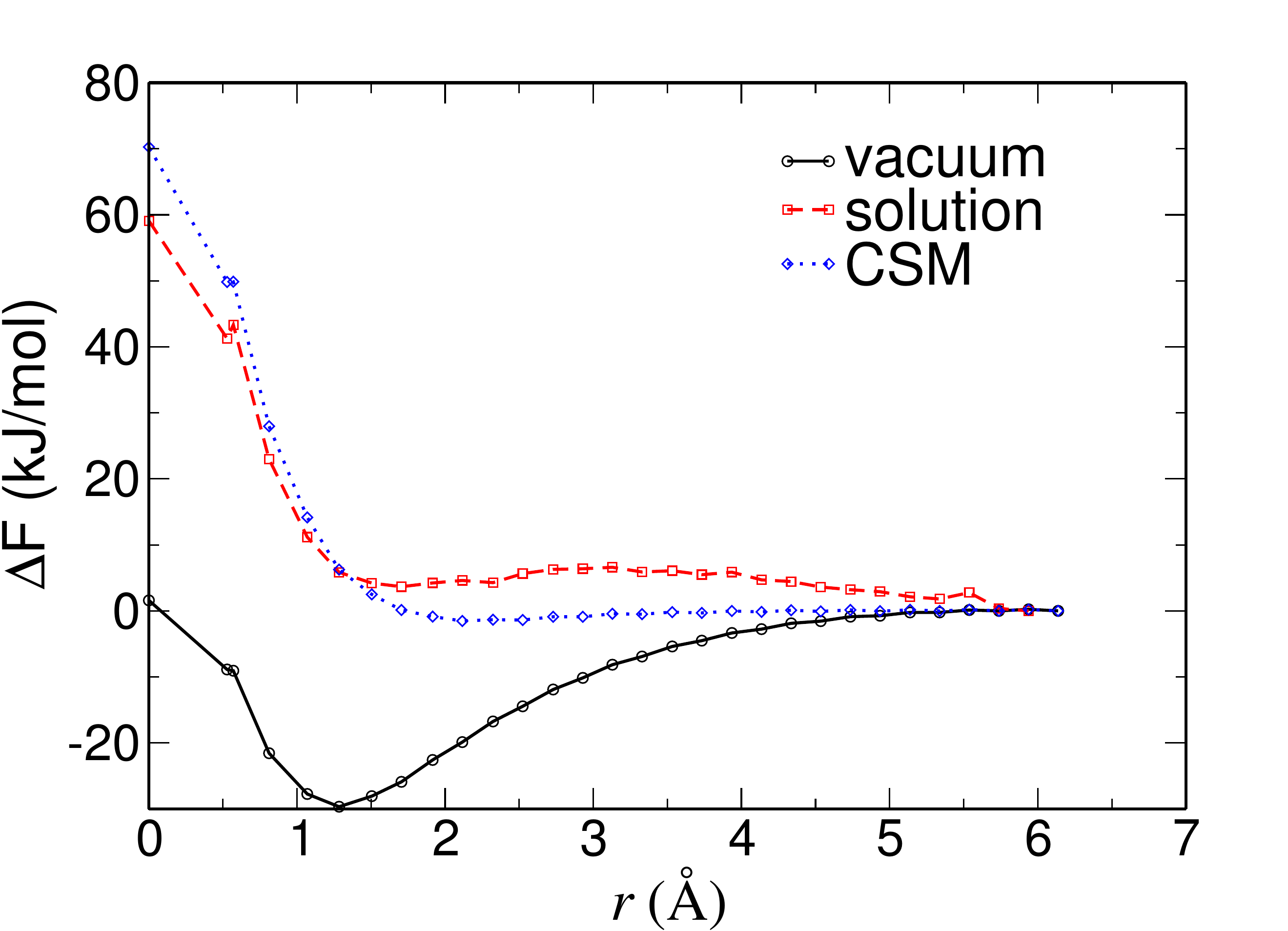}\caption{Potential of mean force of the $\text{S}{}_{\text{N}}2$ reaction
as a function of the reaction coordinate. The calculation in vacuum
is in full black and circles. The free energy in solution obtained
with QM/MDFT are in dashed red with squares. The SFE obtained using
the CSM implemented in GPAW is in dotted blue with diamonds.\label{fig:energy-SN2}}
\end{figure}

We examine the reaction free energy of the symmetrical $\text{S}{}_{\text{N}}2$
reaction between chloromethane and chloride in water. We chose the
same reaction coordinate, $r,$ as many other studies \citep{chandrasekhar_sn2_1984,chandrasekhar_theoretical_1985,huston_hydration_1989,kuechler_quantum_2014,cai_reaction_2019,tirado-rives_qm/mm_2019}
that is the difference between the two carbon chlorine distances $r=\left|d_{\text{C-C\ensuremath{l_{1}}}}-d_{\text{C-C\ensuremath{l_{2}}}}\right|$.
We first run the calculation in the gas phase using GPAW with the
PBE functional and the partial wave basis. The simulation box is
$24\times24\times24\ \textrm{Å}^{3}$. We first identify the transition
state (TS) of the molecule by running calculation on the $\text{\ensuremath{\left[\text{Cl}\lyxmathsym{\textemdash}\text{CH}\ensuremath{_{3}}\lyxmathsym{\textemdash}\text{Cl}\right]}}^{-}$
complex where chlorines and carbon atoms are collinear. Carbon and
chlorine atoms are fixed with two identical carbon chlorine bond length
while the positions of hydrogen atoms are allowed to relax. We vary
the carbon chlorine distances to identify the most stable structure.
In the transition state, the carbon chlorine distance is 2.33~$\textrm{Å}$
and the hydrogen are located on the edges of an equilateral triangle
perpendicular to the $\text{\text{Cl}\textemdash Cl}$ axis, we recover
the expected $D_{3}h$ symmetry for the TS. To compute the energy
profile along the reaction coordinate we elongate the bond between
the carbon and the first chlorine $\text{C}\text{l}_{1}$ and fix
the positions of these two atoms. Other atoms are relaxed with the
constraint of collinearity between carbon and the two chlorines. The
energy profile in vacuum is displayed in figure \ref{fig:energy-SN2},
it exhibits a minimum between the TS and the dissociated state (DS)
which correspond to a so-called ion-dipole complex (IDC). The structure
parameters of TS, IDC and DS are given in table \ref{tab:structure-TS}.
These structures are in overall good agreement with the one reported
by Cai et al \citep{cai_reaction_2019} obtained using M06-2X/6-311++g.
The only major difference is a slightly elongated distance between
carbon and the less bounded chlorine in the IDC.

While the shape of the energy profile in vacuum is correct, the agreement
with previous studies \citep{kuechler_quantum_2014,cai_reaction_2019,tirado-rives_qm/mm_2019}
is not quantitative. We obtain -29.7~kJ.mol$^{-1}$ for the energy
difference between the DS and the IDC. In a recent benchmark, Tirado-Rives
and collaborators have found a stabilisation of the IDC varying between
$-36.0$ and $-51.9$~kJ.mol$^{-1}$ using various methods \citep{tirado-rives_qm/mm_2019}.
We predict an energy barrier, i.e an energy difference between the
TS and the IDC, of $31.3\ $~kJ.mol$^{-1}$ while Bierbaum and coworkers
have found a barrier of $55.2\pm9$~kJ.mol$^{-1}$ using kinetic
analysis \citep{barlow_gas_1988,depuy_gas-phase_1990}. Kuechler and
collaborators have tested a wide variety of methods to compute the
energy barrier and they found an energy difference with respect to
the experimental value up to -19.3~kJ.mol$^{-1}$ \citep{kuechler_quantum_2014}. 

\begin{table}
\centering{}%
\begin{tabular}{|c|c|c|c|c|c|c|}
\hline 
 &  dC-Cl$_{1}$ & dC-Cl$_{2}$ & dC-H & $\angle$HCH & $\angle$HCCl & $r\ \left(\textrm{Å}\right)$\tabularnewline
\hline 
\hline 
TS & 2.33 $\textrm{Å}$ & 2.33 $\textrm{Å}$ & 1.08 $\textrm{Å}$ & 120$\textdegree$ & 90$\textdegree$ & 0\tabularnewline
\hline 
IDC & 1.83 $\textrm{Å}$ & 3.33 $\textrm{Å}$ & 1.09 $\textrm{Å}$ & 110.7$\textdegree$ & 108.1$\textdegree$ & 1.3\tabularnewline
\hline 
DS & 1.79 $\textrm{Å}$ & N/A & 1.09 $\textrm{Å}$ & 110.4$\textdegree$ & 108.4$\textdegree$ & 6.2\tabularnewline
\hline 
\end{tabular}\caption{Structure parameters of the transition state (TS), ion-dipole complex
(IDC) and dissociated state (DS) in the gas phase. \label{tab:structure-TS}}
\end{table}
To study the solvation effect on the symmetric $\text{S}{}_{\text{N}}2$
reaction we coupled the eDFT calculation above with an MDFT description
of the SPC/E solvent. The electronic densities are computed for each
geometry on a regular spatial grid made of $240\times240\times240$
nodes. The same space grid is used for MDFT with 196 possible orientations.
We choose the same set of Lennard-Jones parameters as the one used
by Gao and Xia \citep{gao_two-dimensional_1993} which are reminded
in table \ref{tab:SN2-LJ-parameters}. The eDFT and MDFT program are
run sequentially until a convergence criterion of $10^{-3}$ is reached
for both the relative change in energy for GPAW and in free energy
for MDFT. We apply the usual periodic boundary condition corrections
of charged solutes \citep{kastenholz_computation_2006,kastenholz_computation_2006-1}
and the correction due to the pressure overestimation in HNC \citep{sergiievskyi_fast_2014}.
The solvation free energy computed using eDFT/MDFT is displayed in
figure \ref{fig:energy-SN2}. The minimum corresponding to the IDC
almost vanished while the free energy barrier increases considerably
to 58.7 kJ.mol$^{-1}$. However, this value is still underestimated
compared to the experimental value of 111~kJ.mol$^{-1}$ \citep{albery_methyl_1978}.
The increase of the free energy barrier can be split into two contributions,
the first one being the polarisation of the solute by the solvent.
It can be estimated by comparing the values of the in-vacuo electronic
functional evaluated with the equilibrium electronic densities obtained
in vacuum and in the presence of the solvent. This polarisation contribution
decreases the free energy barrier by 5.9~kJ.mol$^{-1}$. The second
and dominating contribution is the stabilisation by the solvent which
increases the barrier by roughly 29~kJ.mol$^{-1}$. 

The solvation free energy profile was also computed using the CSM
implemented in GPAW \citep{held_simplified_2014}, it is displayed
in figure \ref{fig:energy-SN2}. The free energy barrier is 70.3 kJ.mol$^{-1}$,
a value that is also underestimated with respect to the experimental
one.

The overall agreement of the eDFT/MDFT calculation may seems disappointing
considering that some previous studies were more quantitative, even
with semi-empirical model \citep{kuechler_quantum_2014}. However
this work is the first attempt to self consistently optimise molecular
and electronic functional. It is encouraging that the solvation effects
are well reproduced, at least qualitatively. Moreover, the free energy
profile is consistent with the prediction of the CSM.

There are several avenues to improve the results, first the electronic
functional is clearly not appropriate to reproduce the gas phase predictions,
the MO6-2X functional \citep{zhao_m06_2008} seems more suited for
instance \citep{tirado-rives_qm/mm_2019,cai_reaction_2019}. 

Second, the choice of the Lennard-Jones parameters may not be so innocent.
This is particularly true in this $\text{S}{}_{\text{N}}2$ reaction
where the chlorine atom is described with the same parameters if it
is bonded or in its anionic state. This seems to be natural in QM/MM
studies, for instance in previous studies using the OPLS-AA force
field the parameters for chlorine in halogenoalkane \citep{jorgensen_treatment_2012}
are used for all values of the reaction coordinate. It would probably
be more correct to use a combination of the Lennard-Jones parameters
of the chloride \citep{jensen_halide_2006} and the chlorine in halogenoalkane
depending on the value of the reaction coordinate, this is the object
of current investigation. From the MDFT point of view we recover some
known defects of the HNC functional and the SPC/E model, \textit{i.e.}
a missing bridge term for the former and no explicit treatment of
water polarisability for the latter.

\begin{table}
\centering{}%
\begin{tabular}{|c|c|c|}
\hline 
 & $\sigma\ (\textrm{Å})$ & $\epsilon\ (\text{kJ.mo\ensuremath{l^{-1}}})$\tabularnewline
\hline 
\hline 
Cl & 4.1964  & 0.4682\tabularnewline
\hline 
C & 3.4000 & 0.4184\tabularnewline
\hline 
H & 2.0000  & 0.29288\tabularnewline
\hline 
\end{tabular}\caption{Lennard-Jones parameters for symmetrical S$_{\text{N}}$2 reaction
in water. \label{tab:SN2-LJ-parameters}}
\end{table}

Compared to continuum model a solid advantage of MDFT is its prediction
of the solvent structure at the molecular level. Indeed, the equilibrium
density $\rho(\bm{r},\bm{\Omega})$ contains a lot of information
about the solvation structure. In particular, it is possible to follow
the solvent reorientation during the removal of the nucleofuge. To
do so, we compute the average orientation density $\bm{\bar{\Omega}}$
defined as
\begin{equation}
\bar{\bm{\Omega}}(\bm{r})=\frac{\int\rho(\bm{r},\bm{\Omega})\bm{\Omega}d\bm{\Omega}}{\int\rho(\bm{r},\bm{\Omega})d\bm{\Omega}}.\label{eq:avOmega}
\end{equation}

\begin{figure}
\centering{}\includegraphics[width=0.4\paperwidth]{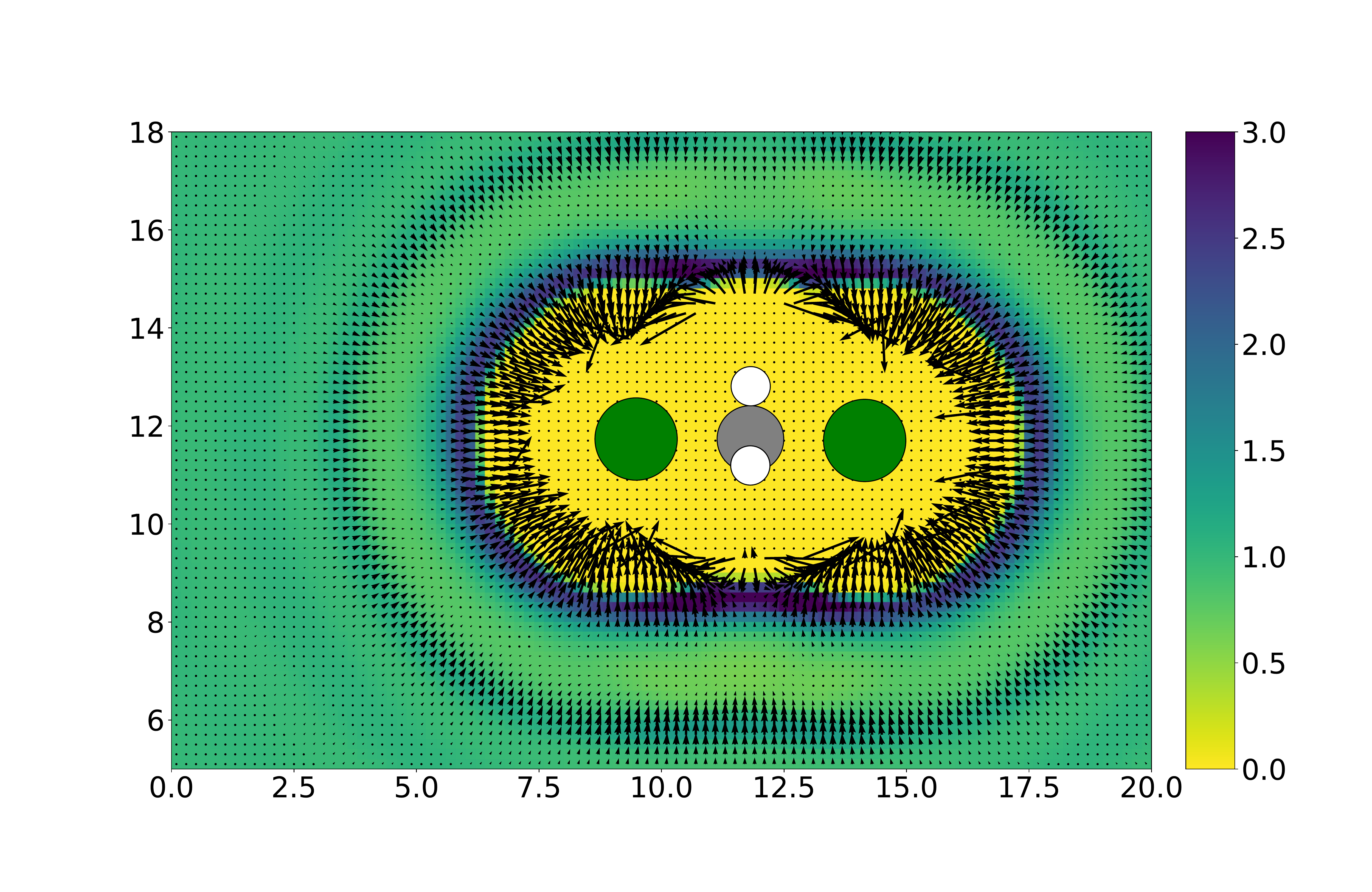}\caption{Average number density of water around the TS in the plane mentioned
in the text. The high density region are in blue, the low density
region are in yellow. The average orientation of the water molecule
as computed by equation \ref{eq:avOmega} are represented by arrows.
Carbon atom is in grey, chlorine atoms are in green and hydrogen atoms
are in white. \label{fig:orientation-TS}}
\end{figure}

\begin{figure}
\centering{}\includegraphics[width=0.4\paperwidth]{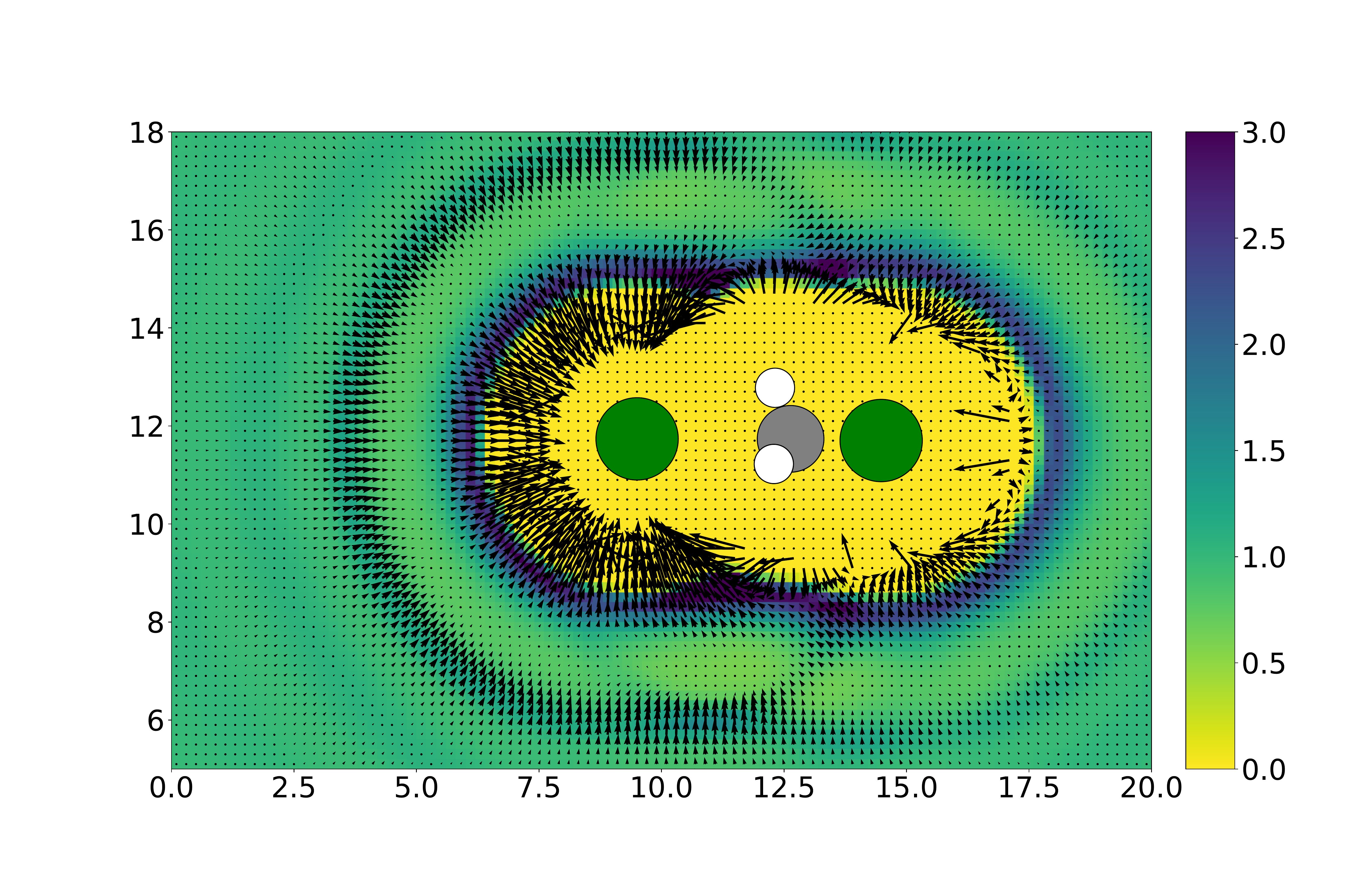}\caption{Same as figure \ref{fig:orientation-TS} for the IDC.\label{fig:orientation-IC}}
\end{figure}

\begin{figure}
\centering{}\includegraphics[width=0.4\paperwidth]{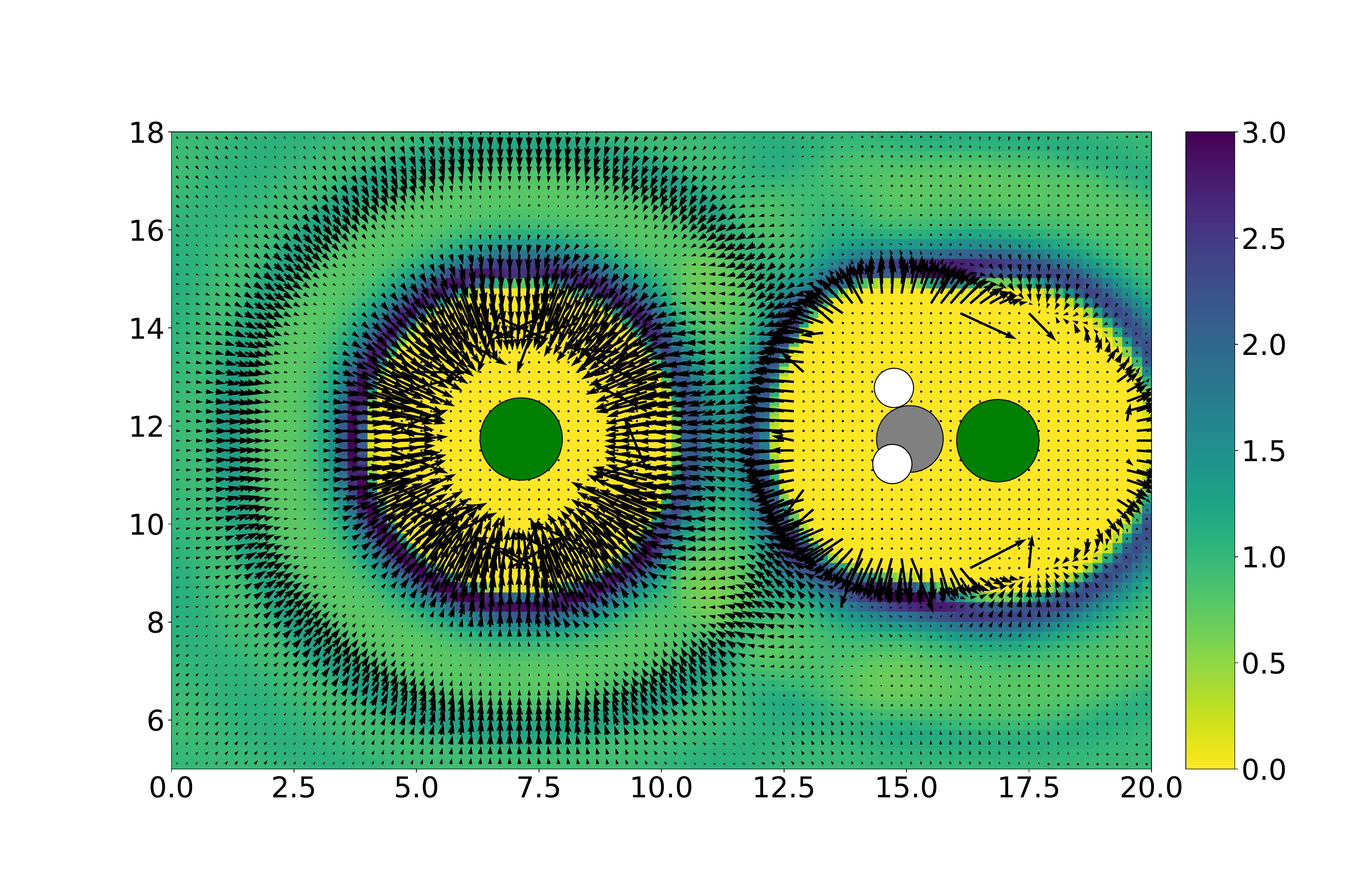}\caption{Same as figure \ref{fig:orientation-TS} for the DS.\label{fig:orientation-diss}}
\end{figure}

The direction of this vector gives the average orientation. Its norm
gives the proportion of the average orientation with respect to other
orientations. The average orientation density and the number density
in a plane containing the carbon, the 2 chlorine and one hydrogen
are displayed in figures \ref{fig:orientation-TS}-\ref{fig:orientation-diss}
for the geometries of table \ref{tab:structure-TS}. The average orientations
are represented by vectors oriented from oxygen towards hydrogens
which length are proportional to $\left\Vert \bm{\bar{\Omega}}\right\Vert $.
To improve the readability of the figure, orientation are depicted
on a grid twice as loose as the one used for calculation. 

Water number densities and average orientation around the TS are symmetrical
with respect to the plane containing the $\text{CH}{}_{3}$ fragment
as displayed in figure \ref{fig:orientation-TS}. Concerning the number
densities, we identify two high density shells separated by a region
where the density is reduced compared to the bulk one. At every location
in the second solvation shell the favoured orientation is the one
with hydrogens pointing towards the solute. This is not surprising
since the solute is globally negative, at this distance it is seen
as a symmetric anion. However, other orientations are not insignificant
as illustrated by the small size of the arrows. In the depletion region,
there are no favoured orientations. In the first solvation shell,
we first mention that the most marked orientation \textit{i.e} the
location denoted by the longest arrows are in the region the closest
to the solute where the number density is almost zero. In this shell,
the favoured orientation at any point are globally pointing towards
the closest chlorine atom, except in a small region located close
to the plane of the bisector of the Cl-Cl bond where the preferred
orientation is pointing outward. Thus, the orientation in the first
solvation shell is consistent with the usual charge picture of the
transition state in which each chlorine is globally anionic and the
CH$_{3}$ fragment is almost neutral \citep{tirado-rives_qm/mm_2019,cai_reaction_2019,kuechler_quantum_2014}. 

The number and average orientation densities around the structure
corresponding to the IDC are displayed in figure \ref{fig:orientation-IC}.
As compared to the TS, the number densities are not drastically modified.
In the first solvation shell, the number density around the leaving
chlorine is increased while the one around the bounded one is decreased.
A similar effect is observed in the second solvation shell but it
is less pronounced. The differences between the average orientations
densities of the IC and the TS are more obvious. In the second solvation
shell, the preferential orientation are still pointing towards the
closest chlorine but the symmetry has clearly been broken. It looks
like the superimposition of two spherical shells centred on each chlorine.
The preferential orientation around the leaving chlorine is even more
pronounced than in the TS while the one around the bounded one almost
already recovered a bulk behaviour with no preferential orientation.
A similar behaviour occurs in the first solvation shell which displays
a decided average orientation around the leaving chlorine while the
preferred orientations towards the bonded chlorine is still present
but drastically reduced. These observations are consistent with an
ionic complex. The two chlorines are anionic but the one the furthest
from the carbon bears a more negative charge than the closest one.

In the dissociated state, displayed in figure \ref{fig:orientation-diss}
the number density and average orientation density follow a similar
trend. The second shell around the chloromethane is no longer visible
neither in number density nor in average orientation density. The
two solvation shells around the leaving chlorine are spherical with
a marked orientation pointing toward the nucleus which is consistent
with an anion. The most interesting feature are observed in the first
solvation shell of the chloromethane where the preferential orientation
in the first solvation shell are pointing outwards the molecule. This
is quite surprising since in the usual point charge model the chloromethane
being described as a dipolar molecule the water molecule close to
chlorine would point towards this atoms. 

\section{Conclusions}

The introduced QM/MDFT approach is well suited to study solvation
problems. It is particularly appropriate to compute solvation free
energies and 3D molecular densities. QM/MDFT is based on the standard
QM/MM approximation where the solute is described by quantum mechanics
and solvent by molecular mechanics using ad~hoc force fields. However,
while the MM system is most often treated using Molecular Dynamics
or Monte-Carlo we employ molecular density functional theory instead.
It is no longer needed to generate extensive trajectory to compute
free energy, this is the direct outcome of a much simpler functional
minimisation. While any variational QM methods could be chosen in
principle, electronic DFT is the most natural to be coupled with classical
DFT \citep{petrosyan_joint_2005,petrosyan_joint_2007}. With this
choice, the solvation free energy can be computed doing a self consistent
optimisation of the electronic and solvent functionals. This could
be done simultaneously but the joint minimisation is done iteratively
in this paper. This self consistent approach account for the mutual
polarisation of the solvent and solute, a phenomena that is disregarded
in previous attempt to couple eDFT and classical DFT \citep{cai_reaction_2019,tang_development_2018}. 

To illustrate the possibilities of QM/MDFT we first studied the solvation
of a quantum water molecule in a solvent of classical water. The water
dipole is enhanced in the liquid as compared to the gas phase. This
is encouraging but the value of the dipole is underestimated with
respect to the experimental measurement. Unfortunately, describing
the solute at the QM level does not fix the known flaws of the MDFT
functional at the HNC level, especially on the solvation structure.
The first peak of the radial distribution between the oxygen of the
solute and the oxygen of the solvent is too broad and the second peak
is misplaced. This calls for bridge functional improvement that are
currently underway.

We then turn to the study of a symmetrical aqueous S$_{\text{N}}$2
reaction between chloromethane and chloride. The free energy profiles
are qualitatively correct, the local minimum observed in gas phase
vanished in the liquid. We were also able to follow in detail the
solvation structure around the reactants along the reaction coordinate.
However, the quantitative agreement between the predicted solvation
free energies and the experimental one is rather disappointing. There
are several room for improvement. First, the electronic functional
should be carefully chosen as different functional can predict energy
barriers differing by several eV. Second, we should reexplore the
corrections to the HNC molecular functional in the light of the QM/MM
problem or better, complement the HNC functional by a well founded
bridge term.

These points surely need to be explored to make the QM/MDFT method
more robust to become a credible alternative to expansive AIMD or
too simplistic continuum solvent models.

\bibliography{QM_MDFT}

\end{document}